\documentclass[doublecol]{epl2} 
\usepackage{amsmath,amsfonts,amssymb,bm}
\usepackage{graphicx,eucal,psfrag,color}% Include figure files
\usepackage{dcolumn}% Align table columns on decimal point
\usepackage{mathbbol}

\def\one{{\mathbb{1}}}
\def\tr{{\rm tr}\; }

\title{Orbital order driven quantum criticality} 
\author{Zohar Nussinov\inst{1} \and Gerardo Ortiz\inst{2}}
\shortauthor{Z. Nussinov \etal}

\institute{                    
  \inst{1} {Department of Physics, Washington University, St.
Louis, MO 63160, USA} \\
  \inst{2} Department of Physics, Indiana University, Bloomington,
IN 47405, USA
}

\abstract{
Charge, spin, and orbital degrees of freedom underlie  the
physics of transition metal compounds. Much work has revealed quantum
critical points associated with spin and charge degrees of freedom in
many of these systems. Here we illustrate that the simplest models that
embody the orbital degrees of freedom - the two- and three-dimensional
quantum  orbital compass models - exhibit an exact quantum critical
behavior on diluted square and cubic lattices (with doping $\delta =1/4$
and $\delta =1/2$ respectively). This raises the possibility of quantum
critical points triggered by the degradation of orbital order upon
doping  (or applying pressure to) such transition  metal systems.
We prove the existence of an orbital spin glass in several related
systems in which the orbital couplings are made non-uniform. Moreover, a
new {\em orbital Larmor precession} (i.e., a periodic change in  the
orbital state) is predicted when  uniaxial pressure is applied.
} 

\begin{document}

\maketitle

{\em{Introduction.}}
%\label{sec1}
The interplay between charge, spin, and orbital degrees of freedom is a
key ingredient underlying the physics of transition metal compounds. The
electronic orbital degrees of freedom often allow for cooperative
effects leading to well defined spatial orderings. Generally, crystal
field effects split the five $n=3$ d-wave orbital wavefunctions into a
triplet ($t_{2g}$ orbitals of the $xy \equiv  (Y_{2,-2} - Y_{2,2})/
\sqrt{2}$,  $yz \equiv  (Y_{2,-1} + Y_{2,1})/ \sqrt{2}$ or  $xz
\equiv(Y_{2,-1} - Y_{2,1})/ \sqrt{2}$ types) and an $e_{g}$ doublet
(spanned by orbitals whose angular dependence is of the
$3z^{2}-r^{2}\equiv Y_{2,0}$ and  $x^2-y^2  \equiv  (Y_{2,-2} +
Y_{2,2})/ \sqrt{2}$ forms). Colossal magnetoresistance (CMR) -- the
sharp decrease of resistivity with magnetic field -- occurs in some of
the compounds that display orbital orders, e.g. the manganite LaMnO$_3$
\cite{bkk} and its derivatives. 
%For the sake of visualization, we briefly depict the nature of the 
%orbital order. 
In LaMnO$_3$, the orbital order is of a simple character: the state of
the single outermost electron in one ion may be that of, say, the $n=3$
d-wave state of the $3z^{2}-r^{2}$ type while it is  $3x^{2}-r^{2}$ on a
neighboring ion and so on in a staggered fashion within each plane. 
Other prominent systems that exhibit various types of orbital order are
found among the vanadates (e.g., V$_2$O$_3$ \cite{v2}, LiVO$_2$
\cite{pen} and LaVO$_3$) \cite{kho} and cuprates (e.g. KCuF$_3$)
\cite{kkh}.  Orbital ordering can be observed via orbital-related 
magnetism and lattice distortions or by resonant X-ray scattering
techniques \cite{mkk}. 

Although there are numerous works on quantum criticality in systems with
various electronic phases \cite{sachdev,vns} there is little prior work
\cite{tnm,sumio,sumio2,karpus} on systems in which the orbital ordering
temperature veered to zero. We wish to motivate the  following question:
{\em If quantum critical points are associated with the degradation of
magnetic/charge/superconducting orders in numerous systems why can these
not appear in orbitally ordered systems?} Such quantum critical points
are associated with the degradation of orbital order. This path has not
been followed despite the success in finding quantum critical points and
novel low-$T$ transitions in spin and other electronic systems. We
demonstrate the existence of exact quantum criticality driven by orbital
order in the simplest of all orbital only models: the $D=2$ and 3
dimensional quantum Orbital Compass Models (OCMs). These systems
rigorously exhibit order in their classical limit
\cite{classicalocm,classicalLONG,Ma}. This order is expected to be
fortified by quantum effects. Indeed, numerically orbital order was
detected in the undiluted $D=2$ OCM \cite{sumio2}.  We will show that,
at a prescribed doping (dilution) of magnitude  $\delta =1/4$,  this
system displays quantum critical correlations.  This quantum critical
phase may be driven away by applying pressure/strain (i.e. varying the
size of the exchange  constants corresponding to different spatial
directions). Similarly, we show that a particular set of anisotropic
couplings of the $D=3$ OCM leads to quantum criticality on a lattice of
doping $\delta=1/2$.
In the particular case that the orbital exchange couplings become random
(i.e., non-uniform), we demonstrate that an {\it orbital spin glass} may
be generated. Furthermore, for any orbital system, when  uniaxial
pressure is applied the phenomenon of {\em orbital Larmor precession}
develops, i.e., the orbital states change periodically in time as a
consequence of the applied pressure.

{\em{Orbital only models. }}
The orbitals in the transition metal oxides are modeled by $S=1/2$
$SU(2)$ degrees of freedom \cite{bkk,new_brink}. These pseudo-spins
appear in the orbital only Jahn-Teller (JT) interactions and
spin-orbital Kugel-Khomskii (KK) Hamiltonians depicting exchange.  The
simplest Hamiltonian linking interactions with the directionality of the
orbitals in space is the  OCM  \cite{new_brink}. The $S=1/2$ anisotropic
$D=2,3$ anisotropic OCMs \cite{NF} on  $N_{\sf{OCM}}= \prod_{\mu}
L_{\mu}$  sites of a square or cubic lattices respectively  are given,
in term of Pauli matrices $\sigma^\mu_j$ with $\mu=x,y$ ($D=2$) or
$\mu=x,y,z$ (for $D=3$), by  
\begin{eqnarray} 
H_{\sf OCM} = -  \sum_{j} J_{\mu} \sigma^{\mu}_{j} \sigma^{\mu}_{j+
\hat{e}_{\mu}} .
\label{ocmeq}
\end{eqnarray}
Interactions involving  the $x$-component of the pseudo-spin $S^x_j= 
\sigma^x_j/2$ occur only along the spatial $x$ direction of the lattice.
Similarly,  spatial direction-dependent spin-exchange interactions
appear for the $y$ (or $z$) components of the spin. In $D=3$, this model
emulates the directional character of the orbital related interactions
triggered by JT effects and the orbital only component of the KK
Hamiltonian \cite{bkk,new_brink,classicalocm,classicalLONG} in materials
with a single electron in a $t_{2g}$ level. 
Varying the strengths of the couplings $\{J_{\mu}\}$ relative to one
another emulates the effects of uniaxial strain and pressures on the
orbital-only component of the orbital-dependent spin-exchange (the KK
Hamiltonian). Similarly, the influence of uniaxial pressure on the JT
interactions may be mimicked by adding a {\em pseudo-spin magnetic
field} linearly coupled to the orbital pseudo-spins. For instance,
uniaxial pressure along the cubic $z$ axis favors orbital states that
have little extent along the $z$ axis. In the case of ${e}_{g}$
orbitals, such a uniaxial pressure favors the planar d$_{x^2-y^2}$
states (represented by $\sigma^{z} = -1$ \cite{bkk,
new_brink,classicalocm, classicalLONG}) vis a vis the cigar-shaped
d$_{3z^{2}-r^{2}}$ state (depicted by $\sigma^{z} = 1$), that is
elongated along the $z$ axis. 

Detailed calculations were performed on orbital systems such as the OCM
\cite{laz}. We will mainly focus on the $D=2$ variant of the OCM. By the
mappings of \cite{NF}, the $D=2$ OCM is dual to $(p+ip)$ superconducting
Josephson junction arrays on a square lattice. Recent theoretical work
suggests similar interactions in $p$-band Mott insulators \cite{p}. 

{\em{The planar orbital compass model.}}
The $D=2$ OCM has  the following symmetries \cite{NO,BN}
%\begin{eqnarray}
$\hat{O}^{\mu}=  \prod_{j \in C_{\mu}} i\sigma_{j}^{\mu} \ \mbox{ , for }
\mu = x,y$ 
%\label{sym2d}
%\end{eqnarray}
with $C_{\mu} \perp \hat{e}_{\mu}$ axis. On a torus, these operators are
defined along toric cycles. As $\hat{O}^{\mu}$ involves
${\cal{O}}(L^{1})$ sites, they constitute $d=1$ symmetries \cite{NO,BN}.
Such symmetries cannot be broken at finite temperatures \cite{BN}. These
are also symmetries of the diluted model that we will describe below. At
the isotropic point $J_{x} = J_{y}$ it has an additional global
reflection symmetry
\begin{eqnarray}
\hat{O}_{\sf reflection} = \prod_{j} e^{i \frac{\pi \sqrt{2}}{4}
(\sigma^{x}_{j} + \sigma^{y}_{j})},
\label{refl}
\end{eqnarray}
which may be broken at finite temperatures \cite{otherrefl}.

\begin{figure}[h]
\includegraphics[width=3.4in]{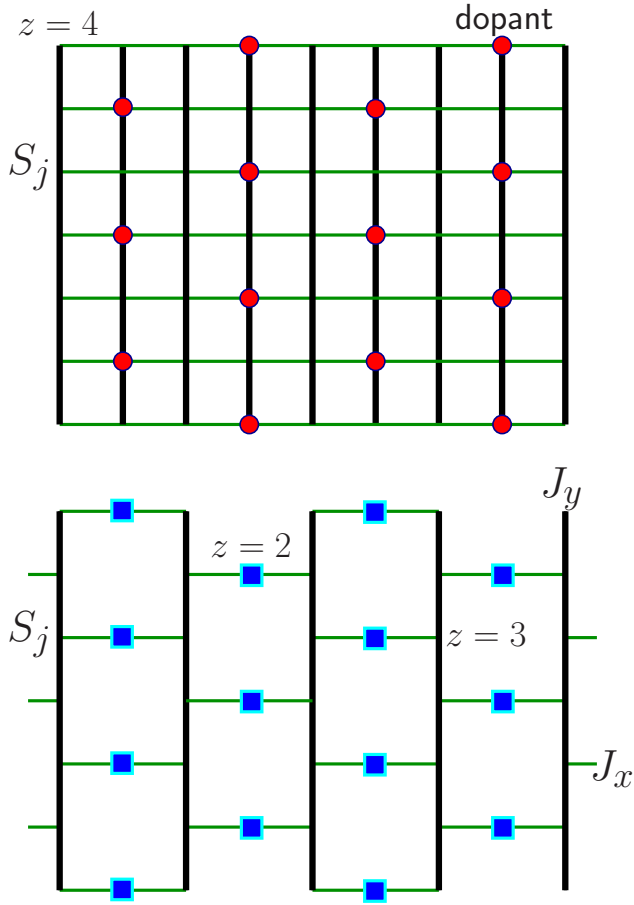}
\caption{Doped $D=2$ orbital compass model. On each vertex of the graph
there is an $S=1/2$ degree of  freedom indicated by a spin-1/2 operator
${S}_{j}$. The top panel represents the undoped orbital compass  model
$(z=4)$ with circles  indicating the dopant sites. Vertical links have
strength $J_y$ while horizontal links have strength $J_x$. The lower
panel displays the resulting doped model where the graph still remains 
two-dimensional (brick-wall topology) but with two types of vertices:
one shares three neighbors ($z=3$) while the other (squares) is only 
connected to two neighbors $(z=2)$. }
\label{fig1}
\end{figure}

As we will prove below, the diluted OCM exhibits a quantum critical
point at the very same location ($J_{x} = J_{y}$) where a finite
temperature transition is expected to occur in the undiluted OCM. It is
natural to anticipate that the ordering temperature drops monotonically
with  doping \cite{sumio2} with, as we will show, the system becoming
quantum critical  for a prescribed dilution of $\delta = 1/4$ (see Fig.
\ref{fig1}).  Away from the isotropic point $J_{x} = J_{y}$, all
symmetries are of the  $\hat{O}^{\mu}$ type. At finite temperatures,
these symmetries cannot be broken and only topological quantum order may
be possible \cite{NO,BN}.  The algebra satisfied by the bond operators
$A_{ij}^{\mu} = \sigma_{i}^{\mu} \sigma_{j}^{\mu}$, for pairs of sites
in which $\vec{j} - \vec{i} = \hat{e}_{\mu}$, is particularly simple:
$\{A^{x}_{ij}, A^{y}_{ik}\} = \{A^{y}_{ik}, A^{x}_{kl}\} = 0, \ \ (j
\neq k \neq i)$, $[A^{\mu}_{ij}, A^{\mu}_{kl}] = 0$,
$(A^{\mu}_{ij})^{2}  =\one$, ${[}A^{\mu}_{ij}, A^{\mu'}_{kl}{]} = 0$, $
(\mu \neq \mu'\ , \ i,j \neq k,l)$. Apart from these algebraic
relations, there are no additional constraints that link the  bonds
$\{A^{\mu}_{ij}\}$.

{\em{A doped orbital compass model.}}
Let us now consider the Hamiltonian of Eq. (\ref{ocmeq}) on the diluted
lattice of Fig. \ref{fig1} (lower panel) with $N=\frac{3}{4}N_{\sf OCM}$
sites, and call this new Hamiltonian $H_{\sf DOCM}$. The lattice of Fig.
\ref{fig1} corresponds to a doped (or diluted) system in which $\delta
=1/4$ of the sites have been replaced by an inert site (dopant).  After
doping, in addition to the symmetries of $\hat{O}^{\mu}$ and 
$\hat{O}_{\sf reflection}$, a new {\it gauge} symmetry emerges 
\begin{eqnarray}
\hat{O}_a=\sigma^x_a , \ \ [H_{\sf DOCM},\hat{O}_a]=0 , 
\label{gaugesym}
\end{eqnarray}
where $a$ represents sites which are two-fold coordinated, i.e. $z=2$
(denoted by squares in Fig. \ref{fig1}). Henceforth we will denote the
set of $z=2$ sites by $\Omega_{2}$.  This gauge symmetry allows to
decompose the total Hilbert space into orthogonal subspaces of equal
dimensionality ${\cal{H}} = \bigoplus_{\ell=1}^{2^{N/3}}
{\cal{H}}_{\ell}$, where $\dim {\cal{H}}_{\ell} =2^{2N/3}$. In each
sector for all states  $|\phi_n \rangle_\ell \in {\cal{H}}_{\ell}$, we
have that  $\sigma^{x}_{a} |\phi_n \rangle_\ell = \eta_{a} |\phi_n
\rangle_\ell \ , \ \eta_a=\pm 1 $. There are $2^{N/3}$ sequences of
eigenvalues of these operators. Thus, each subspace is labeled by a
particular string of $\pm 1$, i.e.  ${\cal{H}}_{\ell} =
{\cal{H}}_{\{\eta_{a}\}_{a=1}^{N/3}}$, with for example
$\{\eta_a\}=+,+,+,-,+,-, \cdots, +,+$.  Each of the Hilbert subspaces
${\cal{H}}_{\ell}$ spans the  $(2N/3)$ spins on the remaining $2N/3$
($z=3$) sites. These sites lie along vertical columns separated by one
another by intervening columns of $z=2$ sites (see Fig. \ref{fig1}). Let
us label the set of all  three-fold coordinates sites $b$ by
$\Omega_{3}$ (the sets $\Omega_2$ and $\Omega_3$ are disjoint
$\Omega_{2} \cap \Omega_{3}  = \{\emptyset\}$). The algebra of the bonds
$\{A_{ij}^{\mu}\}$ in each of the projected subspaces ${\cal{H}}_{\ell}$
is unchanged relative to that defined on the full Hilbert space. To
prove this, let us define the projection operators
\begin{eqnarray}
\hat{P}_\ell=\prod_{a=1}^{N/3} \Big( \frac{\one + \eta_{a}
\sigma^{x}_{a}}{2} \Big) \ , \ \hat{P}_\ell^2=\hat{P}_\ell , \ [H_{\sf
DOCM},\hat{P}_\ell]=0 .
\end{eqnarray}
Within each of the projected subspaces, the bond operators
$\bar{A}_{ij}^{\mu}=\hat{P}_\ell  A_{ij}^{\mu} \hat{P}_\ell$ satisfy
relations identical to those above, i.e. $\{\bar{A}^{x}_{ij},
\bar{A}^{y}_{ik}\} = \{\bar{A}^{y}_{ik}, \bar{A}^{x}_{kl}\} = 0$, $(j
\neq k \neq i)$, $[\bar{A}^{\mu}_{ij}, \bar{A}^{\mu}_{kl}] = 0$,
$\hat{P}_\ell (A^{\mu}_{ij})^{2}\hat{P}_\ell  = \one_\ell$,
${[}\bar{A}^{\mu}_{ij}, \bar{A}^{\mu'}_{kl}{]} = 0$, $(\mu \neq \mu', \
i,j \neq k,l)$. These relations follow directly from the fact that
$[\hat{P}_\ell, A^{x}_{ij}]=0$, since $[\hat{P}_\ell,
\sigma^{x}_{j}]=0$, and $[\hat{P}_\ell, A^{y}_{ij}]=0$, because all
$y$-type bonds  $A_{ij}^{y}$ have their support entirely in the domain
of $z=3$ sites ${i,j} \subset \Omega_{3}$. Consequently, the bond
algebra does not change. 

We now exactly solve the doped OCM (DOCM).  The algebraic relations for
the interaction terms $\bar{A}_{ij}^{\mu}$ defined above are {\em
identical} to those of the interaction terms in the $D=1$ transverse
field Ising model (TFIM). In both instances, the dimension of the
representation of the algebra is the same as are all additional
constraints between the interaction terms. In any given sector
${\cal{H}}_{\ell}$ (which amounts to a particular gauge fix),  all bonds
in the DOCM system are those of decoupled TFIM chains. The Hamiltonian
of a transverse field ($h_j$) Ising chain is
\begin{eqnarray}
H_{\sf TFIM} = - \sum_{j} \Big( h_j  \, \sigma^{x}_{j} + J_{y} \,
\sigma^{y}_{j} \sigma^{y}_{j+ \hat{e}_{y}} \Big).
\label{TFIM}
\end{eqnarray}
On the other hand, the DOCM Hamiltonian of Fig. \ref{fig1} in the
projected subspace, $\bar{H}_{\sf DOCM} \equiv  \hat{P}_\ell H_{\sf
DOCM} \hat{P}_\ell$, is
\begin{eqnarray}
\bar{H}_{\sf DOCM} =  - \sum_{b} \Big( J_{x} \eta_a \; \sigma^{x}_{b}  +
J_{y} \; \sigma^{y}_{b} \sigma^{y}_{b+\hat{e}_{y}} \Big) ,
\label{docm}
\end{eqnarray}
where $a$ denotes the $z=2$ site $a \in \Omega_{2}$  that is a
nearest-neighbor of a given $z=3$ site $b \in \Omega_{3}$ ($a$ lies to
the immediate left or right of $b$ in Fig. \ref{fig1}). $\bar{H}_{\sf
DOCM}$ represents the Hamiltonian of a set of $D=1$ TFIM chains of
length $L$ correlated by the gauge fields $\eta_a$, but otherwise
uncoupled. This is equivalent to say that $\bar{H}_{\sf DOCM}$
represents a {\it single} $D=1$ TFIM chain of length $2N/3$ and
transverse fields $h_b=J_x \eta_a$. The spectrum of a TFIM chain is 
independent of the sign of $h_j$. Given the unitary (and hermitian)
operator $U_\Gamma=\prod_{j\in\Gamma} \sigma^y_j$, where $\Gamma$
denotes an arbitrary set of lattice sites, $U_\Gamma H_{\sf
TFIM}[\{h_j\}]U_\Gamma= H_{\sf TFIM}[\{\bar{h}_j\}]$, where
$\bar{h}_j=-h_j$ if $j \in \Gamma$, and $\bar{h}_j=h_j$, otherwise. 
Thus, all different projected $\bar{H}_{\sf DOCM}$ have exactly the same
spectrum (although their eigenstates are interchanged). As each sector
${\cal{H}}_\ell$ leads to an identical partition function, the resulting
one for the DOCM is
%\begin{eqnarray}
%{\cal Z} &=& \tr_{\cal{H}} e^{-\beta H_{\sf DOCM}} \nonumber \\ 
%&=& \sum_\ell \tr_{{\cal{H}}_\ell} e^{-\beta \bar{H}_{\sf DOCM}} 
%= 2^{N/3} {\cal Z}_{\sf TFIM} ,
%\label{Zc}
%\end{eqnarray}
\begin{eqnarray}
{\cal Z} = \tr_{\!\! \cal{H}} e^{-\beta H_{\sf DOCM}}  = \!\! \sum_\ell
\tr_{\!\! {\cal{H}}_\ell} e^{-\beta \bar{H}_{\sf DOCM}}  = 2^{N/3} {\cal
Z}_{\sf TFIM} ,
\label{Zc}
\end{eqnarray}
where ${\cal Z}_{\sf TFIM}$ is the partition function of the TFIM of Eq.
(\ref{TFIM}) with $2N/3$ sites. If the system  is on a torus of size
$L_{x} \times L_{y}$ we can impose periodic boundary conditions if the
dimensions of the lattice are  such that there is an even number of
sites along a vertical column and a number of sites that is a multiple
of four along the horizontal direction, i.e. $L_{x} \equiv 0 (\mbox{mod}
~4),  \ L_{y} \equiv 0 (\mbox{mod}~ 2)$. In each sector ${\cal{H}}_\ell$
in addition to the algebraic relations defined above, the vertical bonds
will satisfy the additional constraint: $\prod_{b=1}^{L_{y}}
\bar{A}_{b,b+\hat{e}_{y}} =1$. Analogously, the TFIM of  Eq.
(\ref{TFIM}) in which the bonds satisfy the analogous constraint
$\prod_{b=1}^{L_{y}} \sigma^y_b \sigma^y_{b+\hat{e}_{y}} =1$ corresponds
to a system of decoupled chains (formed by the $z=3$ sites)  in each of
which there is a periodic boundary condition (the chain is of length
$L_{y}$). For a system of decoupled $L_{x}/2$ TFIM chains the partition
function is  ${\cal Z}_{\sf TFIM} = (z_{\sf TFIM})^{L_{x}/2}$, where
$z_{\sf TFIM}(L_{y})$ is the partition function of a TFIM chain of
length $L_{y}$. $H_{\sf TFIM}$ can be diagonalized by a Jordan-Wigner
transformation followed by a Bogoliubov transformation, $H_{\sf TFIM}=
\sum_{k} \epsilon_{k} (\gamma^{\dagger}_{k} \gamma^{\;}_{k}-1/2)$,  with
$\gamma^{\;}_{k}$ ($\gamma^{\dagger}_{k}$) the fermionic
annihilation(creation) operator for a fermion (Bogoliubov quasi-particle)
of wavenumber $k$ (the set of allowed wavenumbers is
$\{-\pi,-\pi+\pi/L_y, \cdots,0,\cdots, \pi-\pi/L_y\}$. The quasiparticle
energies of a TFIM chain of length $L_{y}$ are given by $ \epsilon_{k} =
2 \sqrt{J_{x}^{2} + J_{y}^{2} - 2J_{x} J_{y} \cos k}$, and its ground
state energy $E_0=-\sum_k \epsilon_k/2$. Thus, the grand potential per
site is 
\begin{eqnarray}
\!\!\!\!\!\!\!\! \frac{\Omega}{N}&=&\!\!\!-\frac{2E_0}{3L_y} - \!
\frac{k_{B} T \ln 2}{3}
%\nonumber \\ 
-\frac{2k_{B} T}{3L_y}\sum_{k} \ln [ 1+ z e^{- \beta \epsilon_{k}} ],
\end{eqnarray}
with $z=\exp[\beta \xi]$ the fugacity and $\xi$ the chemical potential.
Thus, the DOCM of Fig. \ref{fig1} has a quantum critical point at
$J_{x}=\pm J_{y}$ ($D=2$ Ising universality class). Using our mapping,
in the DOCM, the correlator 
\begin{eqnarray} 
G(\vec{r},t) =
\langle \sigma^{y} (\vec{r}, t)\sigma^{y}(0,0) \rangle_{\cal{H}}
 = \langle \sigma^{y}(\vec{r},t)
\sigma^{y}(0,0) \rangle_{\sf TFIM}.
\end{eqnarray}
Due to the decoupling to TFIM chains, the  last average is zero unless
$\vec{r}$ lies along the $y$ axis. With $c = 2J_{y}$ (the lattice
constant is set to one), for $T=0$, we have at the quantum critical
point $G(\vec{r},0) \sim {(|y|/c)^{-1/4}} \delta_{x,0}$,  and is
manifestly critical. At the quantum critical point for general $T>0$
temperatures, the Fourier transformed correlator is given by
\cite{sachdev} 
\begin{eqnarray}
\tilde{G}(k,\omega) \propto
\frac{\Gamma(\frac{1}{16} - i \frac{\omega + ck}{4 \pi k_{B} T})}
{\Gamma(\frac{15}{16} - i \frac{\omega + ck}{4 \pi k_{B} T})} 
\frac{\Gamma(\frac{1}{16} - i \frac{\omega - ck}
{4 \pi k_{B} T})}{\Gamma(\frac{15}{16} - i \frac{\omega - ck}
{4 \pi k_{B} T})}.
\end{eqnarray}
By our mapping, expressions are similarly available in the paramagnetic
and ordered phases. The regular pattern of doped sites  leads to the
emergence of the gauge symmetry $\hat{O}_a$,  crucial for the exact
solvability of the model. However,  essential to observe lower
dimensional ($d=1$) physics is the presence of the $d=1$  $\hat{O}^\mu$
symmetries \cite{NO}. A modest amount of randomness in the doping  (i.e.
not with a regular pattern) does not invalidate the dimensional
reduction argument, although we can no longer solve the problem exactly.
This conclusion is supported by the numerical simulations of Ref.
\cite{sumio2}. 

Recently, non-uniform structures that may be generated by doping orbital
systems were investigated in Ref. \cite{Kugel}. 

{\em{The doped $D=3$ orbital compass model}. } 
Consider a cubic lattice in which consecutively layered planes (stacked
along the cubic $z$ direction) have the following form:
$ABCBABCBA\cdots$. Here, $B$ planes are as in Fig. \ref{fig1}. Planes of
the $A$ and $C$ types are those in which the corresponding doped columns
of plane $B$ are fully doped. Furthermore, along the columns that are
undoped  in plane $B$, planes $A$ and $C$ are half-doped with  every
other site removed such that there is a relative shift, between the $A$
and $C$ planes, of one lattice constant along the $y$ direction. The
average dopant density on this lattice is $\delta = 1/2$.  Consider the
$D=3$ OCM of Eq. (\ref{ocmeq}) on this lattice. The symmetry of  Eq.
(\ref{gaugesym}) remains intact on all planes of the $B$ type, while 
$\{\sigma^{z}_{u}\}$ for all undoped sites $u$ in the $A$ and $C$ planes
are local symmetries. Replicating the solution of the $D=2$ case, we
find that the $D=3$ DOCM reduces to decoupled TFIM chains in the  $B$
planes as before with a transverse field of strength $h=\sqrt{J_{x}^{2}
+ J_{z}^{2}}$. This is so as along the undoped planar columns of Fig.
\ref{fig1}, each site now feels two transverse  fields along the $x$ and
$z$ directions.  The additional interactions along the $z$ direction
originate from a site in the $A$ {\em or} $C$ planes. No additional $x$
or $y$ interactions appear within the $A$ or $C$ planes. The dispersion
is now given by $\epsilon_{k} = 2 \sqrt{J_{x}^{2}+J_{y}^{2}+J_{z}^{2}- 2
J_{y} \sqrt{J_{x}^{2} + J_{z}^{2}} \cos k}$. This particular lattice
exhibits quantum critical points along the locus $J_{x}^{2} + J_{z}^{2}
= J_{y}^{2}$. As a function of {\em uniaxial pressure} along the $y$
axis,  the point $J_{y}= J \sqrt{2}$ (with $J_{x} = J_{z} \equiv J$) is
quantum critical. For increasing/decreasing uniaxial pressures, and
their influence on the orbital-dependent spin-exchange,  $G(\vec{r},t)$
exhibits the correlations of the ordered/paramagnetic phases of the TFIM
\cite{obvious_half}. 

{\em Uniaxial pressure on Jahn-Teller orbital-only interactions: Orbital
Larmor precession.}
Thus far, we examined the (diluted) anisotropic OCM.  The latter
captures the effects of uniaxial strain and pressure as these apply to
the orbital component of the orbital-dependent spin-exchange
interactions. We now turn to the influence of the uniaxial
strain/pressure on the direct (JT borne) orbital interactions. 
Following our earlier discussion, for a uniaxial pressure/strain along
the $\nu$ space direction, these now generally lead to a Hamiltonian of
the form
\begin{eqnarray}
H_{\sf OCM; ~strain} =H_{\sf OCM} - H_P ,
\label{ocmeqp}
\end{eqnarray}
where the uniaxial pressure is represented by
\begin{eqnarray}
H_P= \sum_{j} P_{\nu} \; \sigma^{\nu}_{j} , 
\label{LR}
\end{eqnarray}
with no summation over $\nu$. For the spatial directions $\nu=x$ or $y$
(here one needs to perform a unitary transformation), the Hamiltonian of
Eq. (\ref{ocmeqp}) can, once again, be exactly solved in the two- and
three-dimensional dilutions considered hitherto. Let us consider
the case $\nu=x$. In this case, the symmetries of Eq. (\ref{gaugesym})
remain manifestly unchanged.  Following our earlier derivation, we
arrive at a simple generalization of Eq. (\ref{docm})
\begin{eqnarray}
\bar{H}_{\sf DOCM; ~strain} = \bar{H}_{\sf DOCM} - \sum_{b} P_{x} \;
\sigma^{x}_{b} -\sum_{a} P_{x} \; \eta_a ,
\label{docm+}
\end{eqnarray}
i.e., a transverse field Ising model with an effective transverse field
$h_b= J_{x} \eta_a+ P_{x}$ while, as before, the Ising exchange constant
$J=J_{y}$. Unlike Eq. (\ref{TFIM}), (\ref{docm+}) does not exhibit a
symmetry of the spectrum under local inversions of the type $\eta_{a}
\to - \eta_{a}$. The general case of uniaxial strain  cannot be exactly
solved and leads to a richer system.  We have seen that the pressure
induced orbital-dependent spin-exchange interactions may, on their own,
lead to a uniaxial pressure tuned quantum critical point. We cannot
trivially solve for the thermodynamics once the direct orbital JT
interactions of the  form of Eqs. (\ref{LR}) and (\ref{docm}) are
included.  The crux is the trace over all $\{\eta_a\}$ that should be 
performed which in effect sums over the partition function of all
subsystems with different transverse fields $h_{b}$.  Hamiltonian
(\ref{docm+}) can be exactly solved within each $\{\eta_a\}$ sector. It
may well be that the quantum critical behavior persists or that, e.g., a
Griffiths-type phase appears in place \cite{sachdev}.

We now comment on a related feature of the diluted OCM when spatially
non-uniform couplings are present. Consider a general system  in which
the (random) exchange constants $\{J_{\mu,j}\}$ vary from bond to bond
(defined by $(j, j + \hat{e}_{\mu})$). Replicating our derivation above,
we find that after applying a uniform  external field that emulates the
effect of uniaxial strain/pressure, we arrive at a transverse field
random Ising model (TFRIM) within each sector of $\{\eta_a\}$, 
\begin{eqnarray}
H_{\sf TFRIM} =  - \sum_{b} \Big( [J_{x,b} \eta_a+ P_{x}]  \;
\sigma^{x}_{b}  + J_{y,b} \; \sigma^{y}_{b} \sigma^{y}_{b+\hat{e}_{y}}
\Big). 
\label{ocmeqpsg}
\end{eqnarray}
In Eq. (\ref{ocmeqpsg}), $P_{x}$ plays the role of a uniform transverse
field while $\{J_{y,j}\}$ are random longitudinal exchange constants. As
the TFRIM exhibits spin glass behavior with well known characteristics,
so may does our system. We may thus predict the appearance of an {\em
orbital spin glass} \cite{FH}.

A fundamental consequence of the form of the uniaxial pressure/strain
induced JT interaction is the appearance of {\em orbital Larmor
precession}. This precession happens even in the simplest bare orbital
system. If only the consequences of uniaxial pressures along a direction
$\nu$  are considered,  then for {\em any} orbital system, the
Hamiltonian $H_P$ of Eq. (\ref{LR}) leads to orbital Larmor precessions.
That is, the orbital states of the electrons change periodically in time
due to the application of pressure
\begin{eqnarray}
\frac{d \vec{\sigma}}{dt} = \gamma \ \vec{\sigma}\times \vec{P} ,
\label{LRE}
\end{eqnarray} 
with $\vec{P} = \sum_{\nu} P_{\nu} \ \hat{e}_{\nu}$, and $\gamma$ the
effective gyromagnetic ratio of the orbital system. The precession times
scale as $|\vec{P}|^{-1}$. In principle, our new orbital precessions
should be observable. 

{\em Possible realizations of orbital ordered triggered quantum
criticality. }  
Our work shows that in orbital systems, upon the application of external
pressure (uniaxial strain) and varying doping, we may attain quantum
criticality associated with the degradation of orbital order.  What
physical systems may display new orbital fluctuation driven quantum
critical points? The Luttinger liquid, a $D=1$ electronic  liquid, and
some variants thereof are prototypical examples of critical systems.
(Indeed, our exact solution relied on a {\it dimensional reduction} to
decoupled chains ensured by the symmetries of these models.) In some
physically probed orbital systems, e.g. \cite{Lee},  an effective
reduction in dimensionality to  precisely such a system occurs due to
the arrangement of orbitals.   KCuF$_{3}$ has a weakly coupled spin
chain structure and offers an example of a system with a $D=1$
(Luttinger liquid) like behavior observed over a broad  range of
temperatures, momenta, and frequencies \cite{Lake}. The orbital only
component of the orbital dependent spin-exchange shares the same
features and anisotropic form as the orbital only interactions
\cite{bkk,kkh}.  The directional character of the orbitals in KCuF$_3$
triggers  a quasi one-dimensional behavior of the spins. Ionic
substitution of KCuF$_3$ (to KCu$_{1-\delta}$Zn$_\delta$F$_3$)
\cite{tnm} revealed a vanishing of the orbital ordering temperature for
a doping of $\delta=1/2$. A reduction in the orbital ordering
temperature occurs as Zn does not have an orbital degree of freedom.
References \cite{sumio,sumio2} noted that this doping fraction is
smaller than the one needed to eradicate order in typical diluted
magnets (e.g. KMn$_{1-\delta}$Mg$_\delta$F$_3$) \cite{DL,breed}; in
typical magnetic systems, the decrease in the ordering temperature  and
its saturation are governed by the percolation threshold (where the
ordering temperature vanishes at the critical dopant concentration of 
$\delta_c=0.69$ for the simple cubic lattice). The faster degradation of
orbital order with doping vis a vis  percolation physics can indeed be
attributed \cite{sumio2}  to the directional character of the orbital
exchange interactions. A loss of orbital order  was also seen in the
bilayered ruthenate Ca$_{3}$Ru$_{2}$O$_{7}$ \cite{karpus} with a
suggested critical pressure of $P \simeq 55$ kbar; this degradation of
orbital order correlates with a metal to insulator transition.  Our work
demonstrates the possibility that {\em quantum criticality is associated
with the vanishing orbital ordering temperature} of systems such as {\rm
KCu$_{1-\delta}$Zn$_\delta$F$_3$} and may be accessed, and its
signatures probed for,  by varying parameters such as the external
pressure and doping. Orbital-only interactions (either borne directly by
the orbital-only Jahn-Teller effect or by the orbital component of the
orbital-dependent spin-exchange) are pertinent to numerous other
systems. These include, for example, the  layered cuprates
(K$_2$CuF$_4$), and manganites (La$_2$MnO$_4$ or La$_2$Mn$_2$O$_7$).
Such systems were discussed recently in \cite{Kugel}.

Note added in proof: Several months after the initial appearance of our
work, Ref.  \cite{Kugel} discussed non-uniform structures that may
appear in doped orbital systems described by orbital models.


\begin{thebibliography}{0}
\bibitem{otherrefl}
In addition to Eq. (\ref{refl}),  the operator
\begin{eqnarray}
U = \exp[\sum_{j} (i \pi/4) \sigma^{z}_{j}] \nonumber
\end{eqnarray}
also effects such an exchange with the difference that $J_{x} \to J_y$
while $J_y \to (-J_{x})$.
\bibitem{obvious_half}
{Obviously, for $\delta =1/2$ we can also have every other site removed
(all sites in which the sum of the $x$, $y$, and $z$ coordinates is
even). That doping corresponds to a set of decoupled Ising spins.
Similarly, if all sites with an even $x$ coordinate are removed,
decoupled $D=2$ OCMs appear (in planes parallel to the $yz$ plane)}.
\end{thebibliography}
\end{document}